\documentclass{aa}
\usepackage{natbib}
\usepackage{graphicx}
\bibpunct{(}{)}{;}{a}{}{,}

\begin{document}
\title{New catalogue of optically visible open clusters and candidates
\thanks { Tables 1a, 1b, 2 and 3  are only available in electronic form at 
http://www.iagusp.usp.br/\~{}wilton/, or at
the CDS via anonymous ftp to ...}
}

\author{W. S. Dias\inst{1}
\and B. S. Alessi\inst{1} 
\and A. Moitinho\inst{2} 
\and J. R. D. L\'epine\inst{1}
}

\offprints{W. S. Dias \email{wilton@iagusp.usp.br}}

\institute{ Universidade de S\~ao Paulo, Dept. de Astronomia, CP 3386,
  S\~ao Paulo 01060-970, Brazil 
  \and Observatorio Astron\'omico
  Nacional, UNAM, Apdo. Postal 877, C.P. 22800, Ensenada B.C.,
  M\'exico }

\date{Received <date>/ Accepted <date>}

\abstract{ 
  We have compiled a new catalogue of open clusters in the
  Galaxy which updates the previous catalogues of \citet{Lynga1987}
  and of \citet{Mermilliod1995} (included in the WEBDA database).  New
  objects and new data, in particular, data on kinematics (proper
  motions) that were not present in the old catalogues, have been
  included.  Virtually all the clusters (1537) presently known were
  included, which represents an increment of about 347 objects
   relative to the \citet{Lynga1987} catalogue.  The catalogue
  is presented in a single table containing all the important data,
  which makes it easy to use.  The catalogue can be accessed on line
  either at \emph{http://www.iagusp.usp.br/\~{}wilton/} or as an
  electronic table which will be made available at the CDS.  
}
\titlerunning{Open Cluster Catalogue}
\authorrunning{Dias et al.}
\maketitle

\section{Introduction}
In this work, we introduce a new catalogue of the open clusters of our
Galaxy. Open clusters have long been recognized as important
tools to investigate the kinematics of star formation regions, aspects
of Galactic structure such as the location of spiral arms, Galactic
dynamics, or even the chemical abundance gradients in the disk.

With the publication of the Hipparcos Catalogue \citep{ESA1997} and
its sub-products, the Tycho \citep{ESA1997} and Tycho2 \citep{Hog2000}
catalogues , and with individual works using CCDs for photometry
and/or astrometry, we have seen a large growth of the available data
on open clusters in a short time.

Among the recent results, we note the discovery of new open clusters
by different authors: \citet{Platais1998} discovered 12 new objects
using Hipparcos data, \citet{Chereul1999} discovered 3 new probable
loose open clusters, and \citet{Dutra2001} discovered 42 objects at
infra-red wavelengths using the 2MASS survey.  Important contributions
were given by \citet{Baumgardt2000} and \citet{Dias2001,Dias2002} who
determined the mean proper motions of more than a hundred clusters,
using the Hipparcos and Tycho2 catalogues, respectively.
\citet{Dias2001,Dias2002} also computed the membership probabilities of the
stars in the cluster fields.  Other recent results are the
determination of the fundamental parameters of 423 clusters by
\citet{Loktin2000} and the discussion of the problem of the
differences in the distances obtained with parallaxes and by
photometric main sequence fitting \citep[see] [ and references
therein]{Pinsonneault1998}.  The latest publications on open  clusters
are divulged in the SCYON electronic newsletter hosted by
the University of Heidelberg
\footnote{http://www.rzuser.uni-heidelberg.de/\~{}s17/scyon/current.html}
in parallel with the WEBDA
database\footnote{http://obswww.unige.ch/webda/}.

Most of the basic data, as well as other results, are included in the
WEBDA database
\citep{Mermilliod1995}, which is the most complete open cluster
database presently available. The WEBDA database includes not only the
data contained in the \citet{Lynga1987} catalogue, which is also a
basic reference much used in the literature, but also provides a huge
amount of additional information. Most of this information is,
however, presented in separate files, available individually for each
cluster.  Also, the database is not updated in what concerns
recently discovered clusters and new designations proposed in the
literature (as discussed in next section). 
 Therefore, the main reasons that prompted us to prepare a
new catalogue, instead of simply adding newly discovered objects, were
the need to have the relevant information in a single file, for
easiness of use, and more important, the fact that the previous
catalogues do not provide the open clusters' proper motions and radial
velocities in a systematic way.

In this work, we inserted the available information on open clusters'
fundamental parameters, kinematics and metalicity in a single file.
We believe that this list will be an important tool for all types of
research on open clusters. Sect.~\ref{sec:cat} describes the contents
of the catalogue and the main reference sources. In
Sect.~\ref{sec:comm} we comment on the new data included.

\section{The catalogue \label{sec:cat}}
In this new open cluster catalogue, we used the previous ones like the
WEBDA, ESO Catalogue \citep{Lauberts1982} and \citet{Lynga1987}
as a starting point. The basic data contained in these
catalogues are coordinates, age, apparent diameter,
 colour excess, and
distance. We inserted new objects, and when available, kinematical
and metalicity data. We made extensive use of the Simbad database and
of the literature to find data on the clusters or on individual stars
of the clusters, to obtain radial velocities and proper motions
averaged over a number of stars.  We do not claim, however, that the
catalogue is the result of a complete survey of all the bibliography
on open clusters.

Our catalogue (Tab.~1a) consists of a single list of
fundamental parameters and kinematical data, with bibliographic
notes. 
The file is self-explanatory and fully documented internally.  The
present version of the catalogue includes information for 1537 open
clusters. For each cluster we list its equatorial coordinates in
J2000.0 and the following parameters, when available: angular apparent
diameter; distance; colour excess; age; mean proper motions and
errors; number of stars used in the proper motion computation and
references; mean radial velocity and error; number of stars used in
the radial velocity determination and references; mean metalicity and
errors; number of stars used in the metalicity determination.  An
identical list (Tab.~1b) is also provided with positions and proper
motions in galactic coordinates.  The full bibliographic references
are given in a separate file (Tab.~2).

In total, 94.7$\%$ of the objects have estimates of their apparent
diameters, and 37$\%$ have distance, $E(B-V)$ and age determinations.
Concerning the data on kinematics, 18$\%$ have their mean proper
motions listed,  12$\%$ their mean radial velocities, and
9$\%$ have both information simultaneously.

Many objects in the list were visually checked in the Digitized Sky
Survey~\footnote{http://archive.stsci.edu/dss/} (DSS) plates, and in
several cases the central coordinates of the clusters were corrected.
This is the case of clusters like \object{Lynga~8}, \object{Stock~12},
\object{Stock~15} and \object{vdB-Hagen~164}, just to mention a few,
that present great differences in position.

Throughout our visual inspection of the DSS plates, there were also
many cases in which no cluster could be found (eg. several Ruprecht,
Collinder and Loden clusters), even in large fields
around their catalogued coordinates.
They were nevertheless kept in the catalogue, but a comment was added.
We shall refer to these ``objects'' as ``non-identified clusters''.
Among these, are the NGC objects flagged as ``non-existent'' in the
\emph{The Revised New General Catalogue
of Non stellar Astronomical Objects} (RNGC)\citep{Sulentic1973}.
On the other hand, some clusters noted as ``non-existent'' in the RNGC
seem to be actual clusters (eg. \object{NGC~2017}, \object{NGC~2609},
\object{NGC~3036}, \object{NGC~5800}, \object{NGC~6115},
\object{NGC~6444}). These have been marked as ``recovered'' in our
catalogue.

A complementary table (Tab.~3) of the clusters with available
photometric data was also built.  Tab.~3 consists of four columns:
cluster name; bands observed with CCD; bands observed with
photomultipiers; bands observed with photographic plates. For each
cluster, only bands with more than ten observed stars are listed. At
the present, the table only lists the UBVRI bands, but it will be
extended to other commonly used photometric systems (eg.  uvby$\beta$,
Geneva, Washington, Vilnius, etc.).  The data table was assembled
using data collected from WEBDA and from searches in the literature.

\section{Comments on new information and new data included \label{sec:comm}}
In this section we comment on some important information given in
the catalogue.

{ \it designations} - An additional remark on the nature of
the open clusters is provided. Among others, we flag the {\it POCR}
\citep[Possible Open Cluster Remnant,][]{Bica2001}. There are 34
objects located at relatively high galactic latitudes ($b\geq
15^{\degr}$) which appear to be  late stages of star cluster
dynamical evolution. The categories also include
possible moving groups like the objects catalogued by Latysev 
and non-identified clusters.

{ \it kinematics} - Recently, many open clusters were investigated and
their mean proper motions \citep{Dias2001,Dias2002,Baumgardt2000}
could be determined. New mean proper motions for 280 objects, and
radial velocities for 182 were inserted in the list.

{ \it Fundamental parameters} - The main source of the fundamental
parameters (reddening, distance and age) was the WEBDA which uses the
information compiled by \citet{Lynga1987}, \citet{Loktin2000},
\citet{Dambis1998} and \citet{Malysheva1997}.
 
All the clusters investigated by \citet{Baumgardt2000} had their
distances estimated from the mean Hipparcos parallaxes of the stars
considered as members.  Recently we investigated 4 open clusters and
determined the mean Hipparcos parallax of stars with membership
probability provided by Tycho2 proper motions \citep{Dias2001}.  The
catalogue includes distances derived from mean parallaxes for Ruprecht
147, Stock 10, vdB-Hagen 23, vdB-Hagen 34, all within 1 kpc.  Also, a
number of parameters from isolated studies were added.
 
{ \it Newly discovered open clusters } - The list includes 191
clusters not present in the previous catalogues. To mention some
cases: \citet{Platais1998} - 12 open clusters were discovered using
Hipparcos data. They are nearby and extended objects; ESO-SC - these
objects (more than 100) were published as probable new open clusters
in the ESO catalogue \citet{Lauberts1982}; Loiano 1- A photometric
study of the surrounding stellar field \citep{Bernabei2001} revealed
that this object lies inside the sky area of a previously undetected
open cluster of intermediate age.  Alessi~1 to 12
are non catalogued objects in the solar vicinity.  Their fundamental
parameters were recently determined showing that they are located at
$d \leq 1 kpc$ \citep{Alessi2002}.

\section{Summary and conclusions}
We have presented a new list of open clusters containing revised data
compiled from old catalogues and from isolated papers recently
published.  This catalogue (Tab.~1a) has been developed mainly to be
an efficient tool for open cluster studies since it presents all the
available basic data (fundamental parameters and kinematics) in a
single easy-to-use list.  The catalogue is regularly updated, and the
latest version is available at
\emph{http://www.iagusp.usp.br/\~{}wilton/}.  An alternative list
(Tab.~1b) with positions and proper motions in galactic coordinates is
also made available. Since it is expected that the catalogue will be
used in the selection of observational targets, an additional table of
open clusters with available photometric data (Tab.~3) is also
provided.  Finally, Tab.~2 includes the references to the data used in
Tabs.~1a and~1b.

In this edition, 1537 objects are given, of which 356 are not given in
the catalogue compiled by \citet{Lynga1987}. The new objects include
191 open cluster published in the literature, and 11 recently
discovered open cluster with fundamental parameters determined by our
group and yet unpublished.

Nearly all the clusters (94.7$\%$) have estimates of their apparent
diameters. Distances, $E(B-V)$ and ages
are listed for 37$\%$ .
Concerning the data on kinematics, 18$\%$ have mean proper
motions determinations,  12$\%$ mean radial velocities, and
9$\%$ have both information simultaneously.
These results point out to the observers that a large effort is still
needed to improve the data on kinematics. Our group is presently
working in this direction.

\begin{acknowledgements}
  We use data from Digitized Sky Survey which were produced at the
  Space Telescope Science Institute under U.S. Government grant NAG
  W-2166.  The images of these surveys are based on photographic data
  obtained using the Oschin Schmidt Telescope on Palomar Mountain and
  the UK Schmidt Telescope. Extensive use has been made of the Simbad
  and WEBDA databases.  This project was supported by FAPESP (grant
  number 99/11781-4).
\end{acknowledgements}

\end{document}